\documentclass[fleqn,twoside]{article}
\usepackage{espcrc2}

\usepackage{graphicx}
\title{\mbox{KASCADE:~Astrophysical results and tests of hadronic
       interaction models}}

\author{A.~Haungs\address[IKUK]{Institut f\"ur Kernphysik, Forschungszentrum
        Karlsruhe, D-76021 Karlsruhe, Germany}\thanks{corresponding 
	author, e-mail: haungs@ik.fzk.de},
        T.~Antoni\address[IKUN]{Institut f\"ur Experimentelle Kernphysik, 
	Universit\"at Karlsruhe, D-76021 Karlsruhe, Germany}, 
        W.D.~Apel\addressmark[IKUK], 
        A.F.~Badea\addressmark[IKUK]\thanks{on leave of absence from 
	{\rm $^b$}},
	K.~Bekk\addressmark[IKUK], 
        A.~Bercuci\address[NIPNE]{National Institute of Physics and Nuclear 
	Engineering, P.O. Box Mg-6, RO-7690 Bucharest, Romania},
        H.~Bl\"umer\addressmark[IKUK]\addressmark[IKUN],
        H.~Bozdog\addressmark[IKUK],
        I.M.~Brancus\addressmark[NIPNE],
        C.~B\"uttner\addressmark[IKUN],
	A.~Chilingarian\address[YPHI]{Cosmic Ray Division, Yerevan
	Physics Institute, Yerevan 36, Armenia},
        K.~Daumiller\addressmark[IKUK],
        P.~Doll\addressmark[IKUK],
	R.~Engel\addressmark[IKUK],
        J.~Engler\addressmark[IKUK], 
        F.~Fe{\ss}ler\addressmark[IKUK],
        H.J.~Gils\addressmark[IKUK],
        R.~Glasstetter\addressmark[IKUN]\thanks{now at Universit\"at Wuppertal, Germany}, 
        D.~Heck\addressmark[IKUK],
        J.R.~H\"orandel\addressmark[IKUN],
        K.-H.~Kampert\addressmark[IKUN]\addressmark[IKUK]{\rm $^\ddag$},
        H.O.~Klages\addressmark[IKUK],
        G.~Maier\addressmark[IKUK]\thanks{now at University Leeds, UK},
        H.J.~Mathes\addressmark[IKUK],
        H.J.~Mayer\addressmark[IKUK], 
        J.~Milke\addressmark[IKUK],
        M.~M\"uller\addressmark[IKUK],
        R.~Obenland\addressmark[IKUK],
        J.~Oehlschl\"ager\addressmark[IKUK],
        S.~Ostapchenko\addressmark[IKUK]\thanks{on leave of absence from Moscow 
	State University, Russia},
        M.~Petcu\addressmark[NIPNE],
        H.~Rebel\addressmark[IKUK],
        A.~Risse\address[SINS]{Soltan Institute for Nuclear Studies,
	PL-90950 Lodz, Poland}, 
        M.~Risse\addressmark[IKUK],
        M.~Roth\addressmark[IKUN], 
        G.~Schatz\addressmark[IKUK],
        H.~Schieler\addressmark[IKUK], 
        J.~Scholz\addressmark[IKUK],
        T.~Thouw\addressmark[IKUK],
        H.~Ulrich\addressmark[IKUK],
	J.~van Buren\addressmark[IKUK],
	A.~Vardanyan\addressmark[YPHI],
        A.~Weindl\addressmark[IKUK], 
        J.~Wochele\addressmark[IKUK], 
        J.~Zabierowski\addressmark[SINS]
       }
       
\begin{document}

\begin{abstract}
KASCADE is a multi-detector setup to get redundant information on
single air shower basis. The information is used to perform multiparameter
analyses to solve the threefold problem of the reconstruction of
(i) the unknown primary energy, (ii) the primary mass, and (iii) 
to quantify the characteristics of the hadronic interactions in the
air-shower development. 
In this talk recent results of the KASCADE data analyses are summarized
concerning cosmic ray anisotropy studies, determination of flux spectra
for different primary mass groups, and approaches to test 
hadronic interaction models.
Neither large scale anisotropies nor point sources were found in the 
KASCADE data set. The energy spectra of the light element 
groups result in a knee-like bending and a steepening above the knee.  
The topology of the individual knee positions shows a dependency on the 
primary particle. Though no hadronic interaction model is fully able to describe
the multi-parameter data of KASCADE consistently, the more 
recent models or improved versions of older models reproduce 
the data better than few years ago. 
\vspace*{-0.5cm}
\vspace{1pc}
\end{abstract}

\maketitle

\section{Introduction}
\begin{figure}[ht]
\vspace*{0.1cm}
\centering
\includegraphics[width=7.5cm]{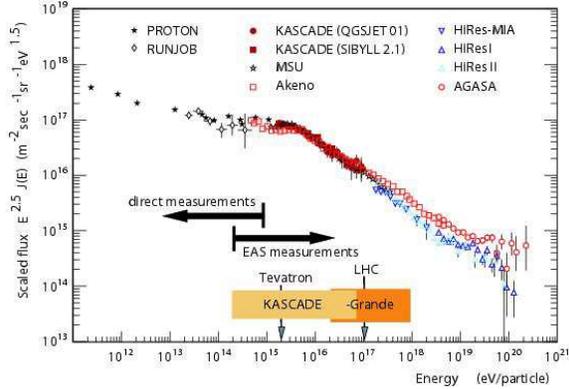}
\vspace*{-1.3cm}
\caption{Primary cosmic ray flux and primary energy range covered by 
KASCADE and its extension KASCADE-Grande.}
\vspace*{-0.5cm}
\label{knee}
\end{figure}

The all-particle energy spectrum of cosmic rays shows a distinctive
feature at few PeV, known as the knee, where the spectral index
changes from $-2.7$ to approximately $-3.1$~(Fig.~\ref{knee}). 
At that energy direct
measurements are presently not possible due to the low flux, but 
indirect measurements observing extensive air showers (EAS) are 
performed. 
Astrophysical scenarios like the change of the acceleration 
mechanisms at the cosmic ray sources 
(supernova remnants, pulsars, etc.) 
or effects of the transport mechanisms inside the Galaxy (diffusion 
with escape probabilities) are conceivable for the origin of the knee
as well as particle physics reasons like a new kind of hadronic 
interaction inside the atmosphere or during the transport through 
the interstellar medium. 

Despite EAS measurements with many experimental setups in the 
last five decades the origin of 
the kink is still not clear, as the disentanglement of the threefold 
problem of estimate of energy and mass plus the understanding of
the air-shower development in the Earth's atmosphere remains an 
experimental challenge. For a detailed discussion of the subject
see a recent review~\cite{rpp}. 

The multi-detector system KASCADE (KArlsruhe Shower Core and Array 
DEtector)~\cite{kas} approaches this challenge by 
measuring as much as possible redundant information from
each single air-shower event. 
The multi-detector arrangement allows 
to measure the total electron and muon numbers of the shower 
separately using an array of shielded and unshielded detectors 
at the same place. 
Additionally muon densities at further
three muon energy thresholds 
and the hadronic core of the 
shower by an iron sampling calorimeter are measured. 

In the following we present the main results of KASCADE,
in particular the reconstruction of energy spectra of single primary mass 
groups and approaches of correlation analyses to test the hadronic 
interaction models and to find constraints for the improvement and 
development of the next generation of the models. 
These tests provide complementary information to the data of present
accelerator experiments, as air-shower data are sensitive to 
higher energy interactions and to a different (extreme forward direction) 
kinematic region. 

\section{The KASCADE experiment}

The KASCADE experiment~\cite{kas} measures 
showers in a primary energy range from $100\,$TeV to $80\,$PeV 
and provides 
multi-parameter measurements on a large number of observables concerning 
\begin{figure}[h]
\vspace*{0.1cm}
\begin{center}
\includegraphics[width=5.3cm]{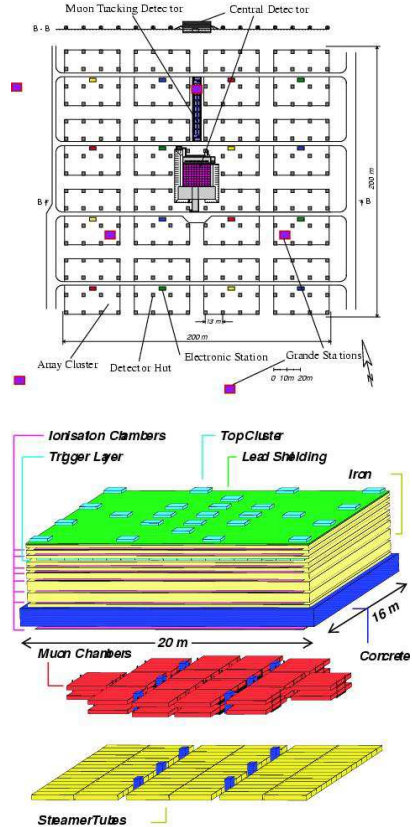}
\end{center}
\vspace*{-1.0cm}
\caption{The main detector components of the KASCADE experiment: (the 16
clusters of) Field Array, Muon Tracking Detector and Central Detector. 
The location of some stations of the Grande array is also displayed.
The lower part of the figure shows a sketch of the Central Detector with 
its different detection set-ups.}
\vspace*{-0.5cm}
\label{KASCADE}
\end{figure}
electrons, muons at 4 energy thresholds, and hadrons.
The main detector components of KASCADE are the Field Array, the Central
Detector, and the Muon Tracking Detector (Fig.~\ref{KASCADE}). 

The Field Array measures the 
total electron and muon numbers ($E_\mu>240\,$MeV)
of the shower separately using an array of 252 detector stations 
containing shielded and unshielded detectors at the same place 
in a grid of $200 \times 200\,$m$^2$.
The excellent time resolution of these detectors allows also decent 
investigations of the arrival directions of the showers in searching
large scale anisotropies and, if exist, cosmic ray point sources. 

The Muon Tracking Detector measures the incidence angles 
of muons ($E_\mu > 800\,$MeV) 
relative to the shower arrival direction. 

The hadronic core of the shower is measured by a $300\,$m$^2$ 
iron sampling calorimeter installed at the KASCADE Central Detector: 
Three other components - trigger plane (serves also as timing
facility), multiwire proportional chambers (MWPC), 
and limited streamer tubes (LST) - offer additional 
valuable information on the penetrating muonic component 
at $490\,$MeV and $2.4\,$GeV energy thresholds. 

The redundant information of the showers measured by the 
Central Detector and the Muon Tracking Detector
is prevailingly used for tests and improvements 
of the hadronic interaction models unavoidably needed for the 
interpretation of air shower data. 

\section{KASCADE results}
First, some results which are obtained nearly
independently of influences of the simulations, 
in particular studies of anisotropies of cosmic rays and gamma 
ray search will be discussed.
Second, results of the main analyses of KASCADE are reported, 
i.e.~the unfolding of the 
two-dimensional electron number to muon number size spectrum in 
energy spectra of five different mass groups and the dependence
of the results on the hadronic interaction models.
Finally, specific tests of the hadronic interaction models using 
data of the Central Detector and the Muon Tracking Detector 
of KASCADE are discussed. 

\subsection{Search for anisotropies and point sources}
Investigations of anisotropies in the arrival directions of the 
cosmic rays give additional information on the cosmic ray origin 
and of their propagation.  
Depending on the 
\begin{figure}[t]
\vspace*{0.1cm}
\centering
\includegraphics[width=5.7cm]{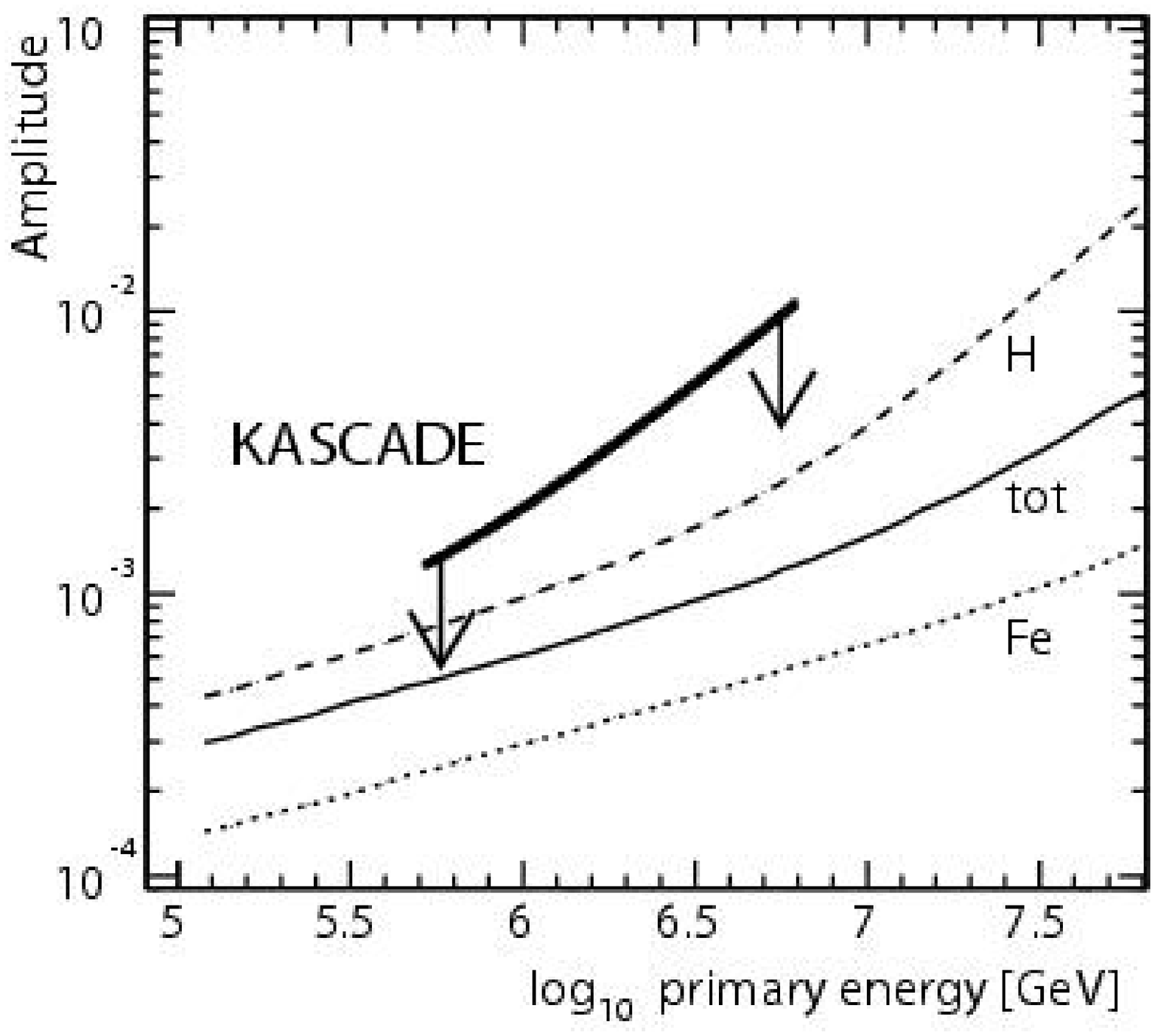}
\includegraphics[width=5.7cm]{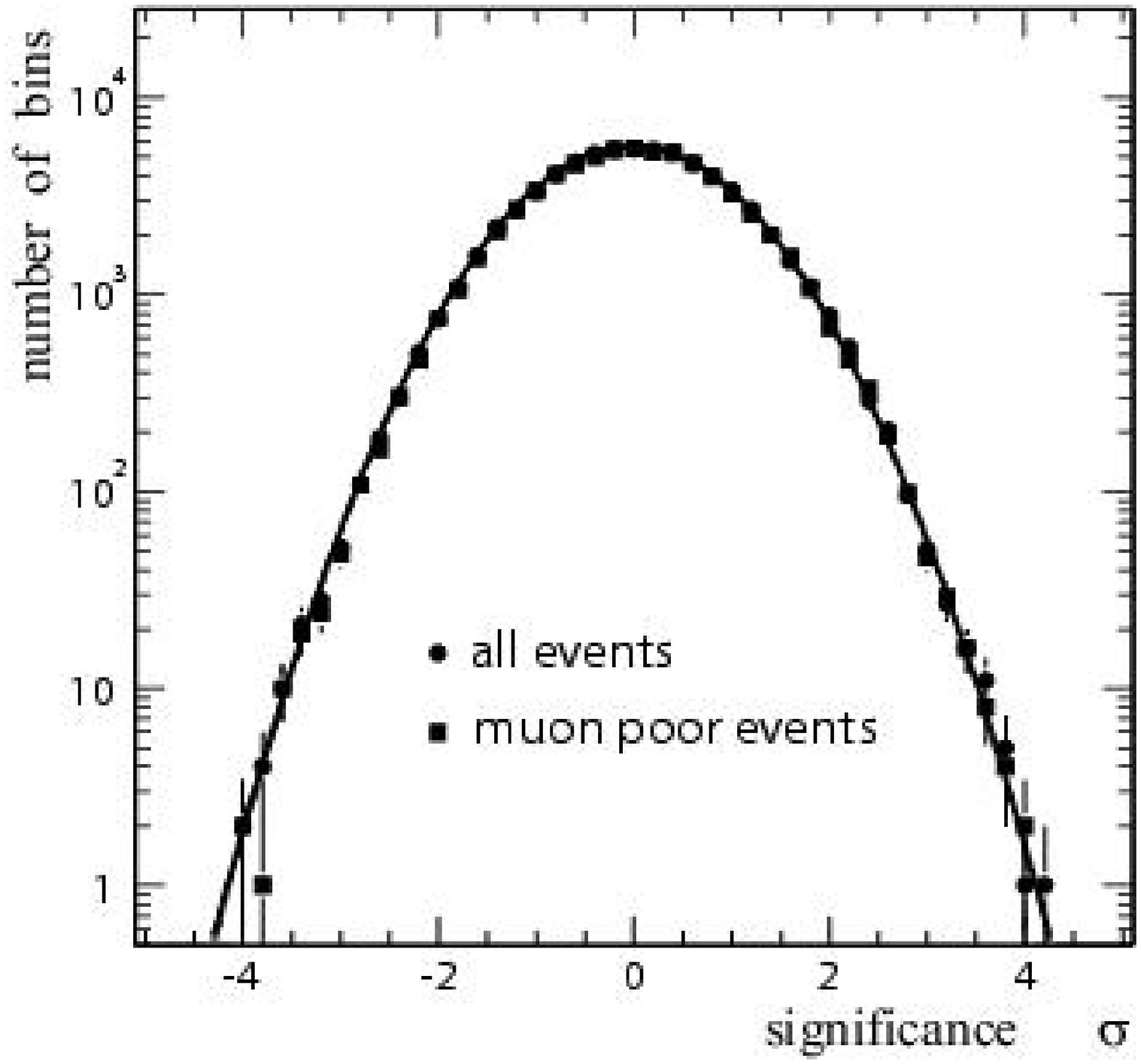}
\vspace*{-.8cm}
\caption{Upper part: Rayleigh amplitude of the harmonic analyses 
of the KASCADE data~\cite{antoni:lsanis} (limit on a 95\% 
confidence level) compared to theory predictions~\cite{candia}. 
Lower part: Significance distributions for searching
point sources on the sky map seen by the KASCADE 
experiment~\cite{antoni-ps}.}
\vspace*{-0.5cm}
\label{aniso}
\end{figure} 
model of the origin of the knee one expects large-scale anisotropies 
on a scale of $10^{-4}$ to $10^{-2}$ in the energy region of the knee
and depending on the assumed structure of the galactic magnetic field.
For example in Fig.~\ref{aniso} (upper panel) the predictions 
from calculations of Candia et al.~\cite{candia}~are compared with 
the limits of anisotropy given by KASCADE 
results~\cite{antoni:lsanis}. 
The KASCADE limits were obtained by investigations of 
the Rayleigh amplitudes and phases of the first harmonics. 
Taking into account possible nearby sources of galactic cosmic 
rays like the Vela Supernova 
remnant~\cite{ptuskin} the limits of KASCADE already exclude 
particular model predictions. 
\begin{figure}[ht]
\vspace*{0.1cm}
\includegraphics[width=6.8cm]{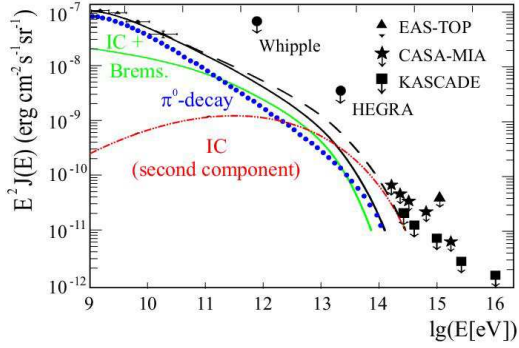}
\vspace*{-0.8cm}
\caption{Derived limits on the gamma flux compared to limits of 
other experiments and theoretical spectra from~\cite{aharonian}.}
\vspace*{-0.5cm}
\label{gamma}
\end{figure}
But for a complete picture the investigations have to be performed 
with air shower samples of the different mass groups
which need a higher statistical accuracy in measurements. 

Some interest for looking to point sources in the KASCADE data sample 
arises from the possibility of unknown near-by sources, where the 
deflection of the charged cosmic rays would be small or by sources 
emitting neutral particles like high-energy gammas or neutrons. 
Due to their small decay lengths the latter ones are of interest 
for near-by sources only. Fig.~\ref{aniso} (lower panel) shows the 
distribution of significances for a deviation of the flux from the 
expected background for all bins of the visible sky of KASCADE. 
Shown are  the distributions 
for the full sample of air showers as well as for a sample of 
"muon-poor" showers which is a sample with an enhanced number 
of candidates of $\gamma$-ray
induced events. No significant excess was found in both
samples~\cite{antoni-ps}.  

With a similarly obtained sample of enhanced gamma candidates an analyses 
is performed to estimate limits~\cite{schatz} on the diffuse high-energy 
gamma ray flux. Fig.~\ref{gamma} shows the flux compared to obtained limits 
from other experiments and theoretical spectra from~\cite{aharonian}.
By increasing the statistics and/or improving the selection procedures the
KASCADE results will reach the sensitivity to proof such theoretical 
calculations.

\subsection{Energy spectra of individual mass groups}
\label{unfolding}
\begin{figure}[ht]
\vspace*{0.1cm}
\centering
\includegraphics[width=7.3cm]{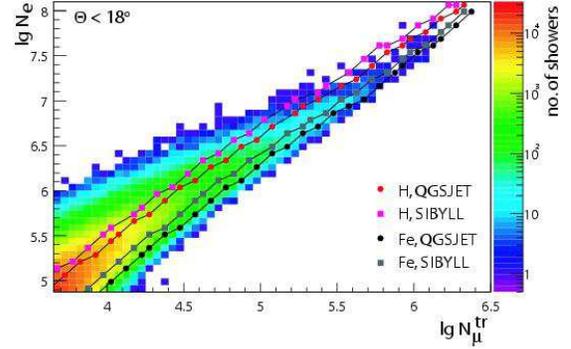}
\vspace*{-.8cm}
\caption{Two dimensional electron ($N_e$)~vs.~muon 
($N_\mu^{\rm tr}\,=\,$number of muons in 40-200m core distance) 
number spectrum measured by the KASCADE array. 
The lines display the most probable values for proton
and iron primaries obtained by CORSIKA simulations employing 
different hadronic interaction models.} 
\label{data}
\vspace*{-0.5cm}
\end{figure}

The content of each cell of the two-dimensional spectrum of 
electron number~vs.~muon number (Fig.~\ref{data}) is the sum of
contributions from the individual primary elements.
Hence the inverse problem 
{\small $g(y) = \int{K(y,x)p(x)dx}$} 
with {\small $y=(N_e,N_\mu^{\rm tr})$} and {\small $x=(E,A)$} 
has to be solved.
This problem results
in a system of coupled Fredholm integral equations of the form \\
{\small  $\frac{dJ}{d\,\lg N_e\,\,d\,\lg N_\mu^{tr}} = $ \\
 \hspace*{1cm} $ \sum_A \int\limits_{-\infty}^{+\infty} \frac{d\,J_A}{d\,\lg E} 
  \cdot 
  p_A(\lg N_e\, , \,\lg N_\mu^{tr}\, \mid \, \lg E)
  \cdot 
  d\, \lg E $ } \\
where the probability $p_A$  
is a further integral with the kernel function 
{\small $k_A = r_A \cdot \epsilon_A \cdot s_A$}
factorized into three parts. The quantity $r_A$ describes 
the shower fluctuations, 
i.e. the distribution of electron and muon number for given 
primary energy and mass. The quantity $\epsilon_A$ describes 
the trigger efficiency of the experiment, 
and $s_A$ describes the reconstruction
probabilities, i.e. the distribution of reconstructed $N_e$ and 
$N_\mu^{\rm tr}$ for given true numbers of electrons and muons.
\begin{figure}[ht]
\vspace*{0.1cm}
\begin{center}
\includegraphics[width=6.5cm]{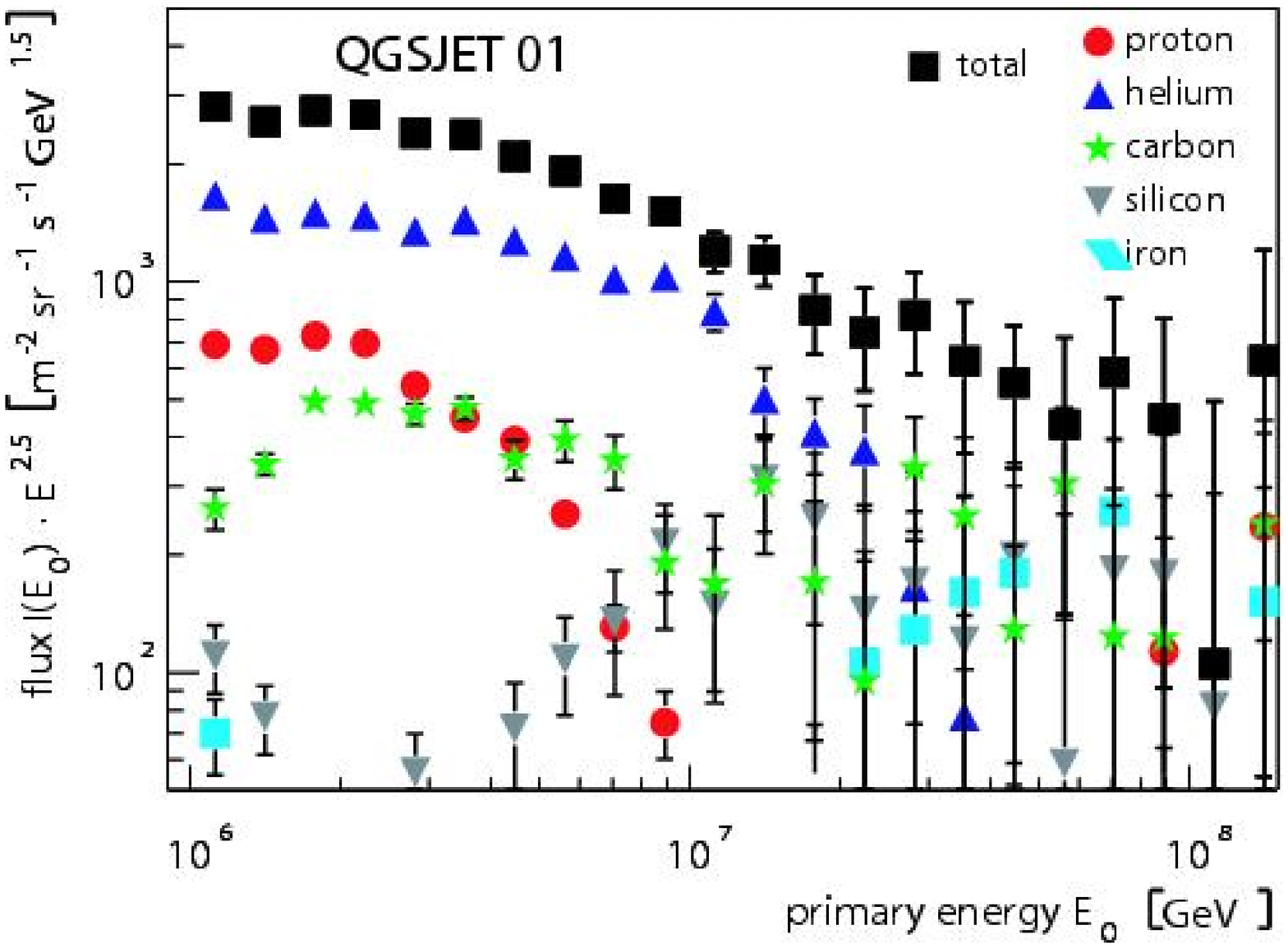}
\includegraphics[width=6.5cm]{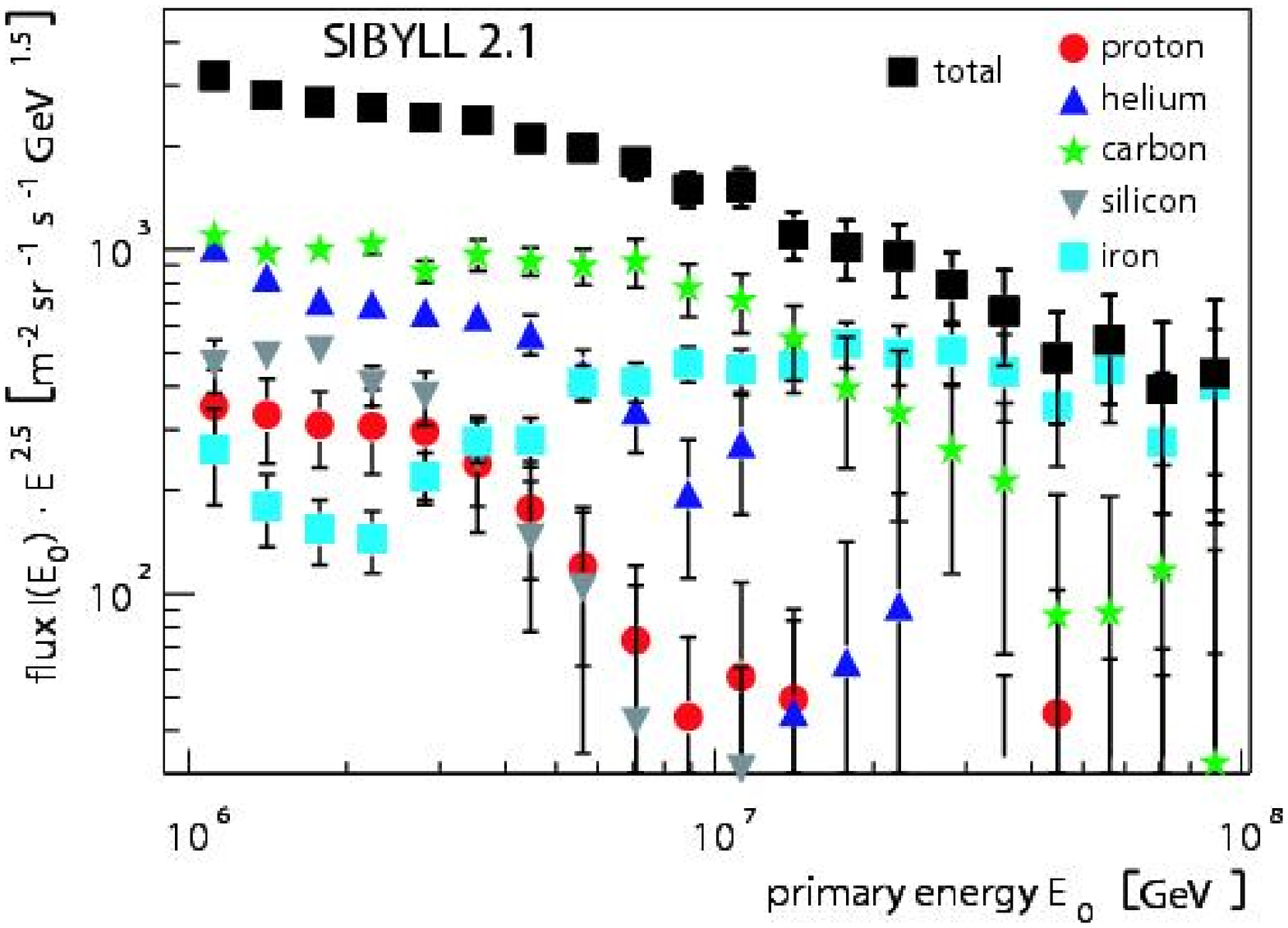}
\end{center}
\vspace*{-1.1cm}
\caption{Result of the unfolding procedure. Upper part: based
on QGSJET$\,01$; Lower part: based on SIBYLL$\,2.1$.}
\vspace*{-0.5cm}
\label{spectra}
\end{figure}
The probabilities $p_A$ are obtained by parameterizations of 
Monte Carlo simulations for fixed energies using a
moderate thinning procedure as well as fully simulated showers 
as input of the detector simulations. 

The application of the unfolding procedure to the data 
is performed on basis of 
two different hadronic interaction models (QGSJET$\,01$~\cite{qgs}, 
SIBYLL$\,2.1$~\cite{sib}) as options embedded in 
CORSIKA~\cite{cors} for 
the reconstruction of the kernel functions~\cite{roth}. 
By applying the above described procedures to the experimental 
data energy spectra are obtained as displayed in 
Fig.~\ref{spectra}.

Knee like features are clearly visible in the all particle spectrum,
which is the sum of the unfolded single mass group spectra, 
as well as in the spectra of primary proton and helium.
This demonstrates that the elemental composition of cosmic
rays is dominated by light components below the knee and dominated
by a heavy component above the knee feature. Thus the knee feature 
originates from a decreasing flux of the light primary particles.
This observation corroborates results of the analysis of muon 
density measurements
at KASCADE~\cite{kas-muon}, which are performed independently 
of the unfolding procedure. 

\subsection{Inaccuracies of hadronic interaction models}
Comparing the unfolding results based on the two different hadronic 
interaction models, the model dependence when interpreting the data
is obvious. 
Modeling the hadronic 
interactions underlies assumptions from particle physics theory and 
extrapolations resulting in large uncertainties, which are reflected 
by the discrepancies of the results presented here. 
The most prominent difference lies in the larger contribution of heavier 
primaries in case of the SIBYLL model, especially at high energies.
To understand the differences between the results based on the two different 
models and to judge the validity of the models in general 
detailed investigations of the results are performed.
In Fig.~\ref{data} the predictions of the $N_e$ and $N_\mu^{tr}$ 
correlation for the 
two models are given in case of proton and iron primaries. 
It is remarkable that all four lines have a more or less parallel 
slope which is different from the data distribution. There, the knee is 
visible as kink to a flatter $N_e$-$N_\mu^{tr}$ 
dependence above $N_\mu^{tr} \approx 4.2\,$. The heavier primary contribution 
on the results based on the SIBYLL model is due to predictions of
a larger ratio of muon to electron number for all primaries. 
Comparing the residuals of the unfolded two dimensional distributions for 
the different models with the initial data set we conclude~\cite{hep-aachen} 
that at lower energies the SIBYLL model and at higher energies 
the QGSJET model are able to describe 
the correlation consistently, but none of the present models
gives a contenting description of the whole data set. 
Also a preliminary analysis using the FLUKA~\cite{fluka} code instead of
GHEISHA~\cite{gheisha} as low energy interaction model shows no conspicuous
improvement of the situation.

Crucial parameters in the modeling of hadronic 
interaction models which can be responsible for these inconsistencies 
are the total nucleus-air cross-section and the parts of the inelastic and 
diffractive cross sections leading to shifts of the position of the 
shower maximum in the atmosphere, and therefore to a change of the muon
and electron numbers as well as to their correlation on single air 
shower basis. 
The multiplicity of the pion generation at all energies at 
the hadronic interactions during the air shower development is also
a 'semi-free' parameter in the air-shower modeling as accelerator data
have still large uncertainties.

\subsection{Tests with hadronic observables}

Arbitrary changes of free parameters 
in the interaction models will change the 
correlation of all shower parameters. Tests using KASCADE observables,
which are measured independently of such used in the unfolding procedure,
may give further constraints, e.g.~by investigating correlations of 
the hadronic shower component with electron or muon numbers. 
The aim is to provide 
hints for the model builder groups how the parameters (and the theory) 
should be modified in order to describe all the data consistently.

The applied method here is to evaluate the measured data relative to
simulations (including the detector response) 
of proton and iron primaries. The measurements have to 
lie between these extreme values, otherwise the simulations cannot
describe this specific observable correlation. Direct comparisons
between data and simulations are not possible due to the unknown
composition of the primary particles generating the air showers.
 
\begin{figure}[ht]
\vspace*{0.1cm}
\centering
\includegraphics[width=7.cm]{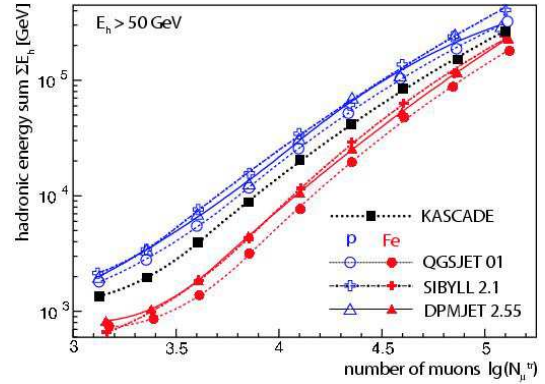}
\vspace*{-0.8cm}
\caption{Correlation between the reconstructed hadronic energy sum 
and the number of muons.}
\label{seh-nmu}
\vspace*{-0.5cm}
\end{figure}

These kind of tests are performed for a large set of interaction models 
employed in the simulation package CORSIKA~\cite{cors}. 
In ref.~\cite{JphysG-hc} first results of these tests were published
investigating the models VENUS~4.12~\cite{venus},
SIBYLL~1.6~\cite{sib16}, and QGSJET~98~\cite{qgs}.
The general conclusion was that QGSJET
described the data best at that time, whereas
strong hints could be given that SIBYLL~1.6 generates too few
muons. These result triggered improvements of the model 
leading to the newer version SIBYLL~2.1~\cite{sib}. 
Later~\cite{milke}, NEXUS~2~\cite{nexus}, SIBYLL~2.1 and 
QGSJET~01~\cite{heckicrc} were
investigated with the result that the differences between the models
got smaller. Whereas QGSJET~01 and SIBYLL~2.1 can describe now the 
KASCADE hadronic observables, NEXUS calculations predict too little hadronic
energy at observation level.   
Present investigations comparing the data with DPMJET~2.55~\cite{dpmjet}, 
QGSJET~01, and SIBYLL~2.1 
confirm that these three models can describe the hadronic observables
and their correlations with the electron and muon component 
within the sensitivity of the KASCADE experiment (Fig.~\ref{seh-nmu}), 
at least in the energy range below 10 PeV. 

The tests will be continued with new models, e.g.~QGSJET II~\cite{qgsII}, 
and with better statistical accuracy at higher primary energies.
Additionally another approach will be followed: To change specific 
parameters in one certain model (QGSJET~01)
and to investigate the change of the correlations of measurable 
observables~\cite{jrh04} and compare them with the data.

A different window to the hadron-muon correlation is the investigation
of the KASCADE trigger rates~\cite{tr-rate}. Here, in particular
primaries of lower energies and their behavior in the atmosphere is 
approached. By this, indications for an underestimate of the
non-diffractive inelastic cross-section in the models are given.
  
\subsection{Tests with muon densities}
\begin{figure}[ht]
\vspace*{0.1cm}
\centering
\includegraphics[width=5.7cm]{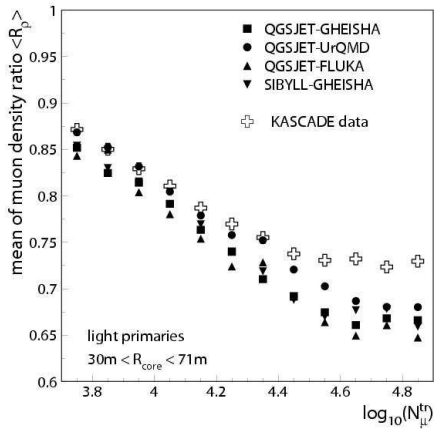}
\vspace*{-0.8cm}
\caption{Distributions of the muon density ratio for KASCADE data 
in comparison with predictions for various model combinations.}
\label{muratio}
\vspace*{-0.5cm}
\end{figure}

The tests described in last sections are sensitive to the high-energy 
interaction models used in EAS simulations. Measurements of muon
densities for different energy thresholds, here 2.4 GeV and 490 MeV
detected by the KASCADE Central Detector are more sensitive to 
low-energy interaction models like GHEISHA~\cite{gheisha}, 
UrQMD~1.1~\cite{urqmd}, or FLUKA~\cite{fluka} 
(see also ref.~\cite{heck}).
The ratio of muon densities measured at fixed core distances 
are determined by the muon energy spectrum in air-showers, 
which is a different approach to test the models
than investigating the total number of muons~\cite{haungs03}.
Fig.~\ref{muratio} displays the ratio of the muon densities
for two thresholds in dependence of the muon
number (truncated muon number) measured by the KASCADE array
($E_\mu^{thr}=240\,$MeV). The range of $N_\mu^{tr}$ corresponds to
the primary energy range of $10^6-10^7\,$GeV. 
Compared is the distribution of the data with
predictions (including full detector response and reconstruction
procedures) for various high-energy and low energy model combinations.
None of the models can reproduce fully the measurements, especially at
higher muon number, i.e.~higher primary energy, but 
the behavior of the UrQMD model seems to be more consistent
as FLUKA or GHEISHA. 
Next generation of CORSIKA will include QGSJET~II~\cite{qgsII} as a new
model, which show in first calculations a significantly different 
behavior of the muon component.

\subsection{Tests with muon pseudorapidities}
\begin{figure}[ht]
\vspace*{0.1cm}
\centering
\includegraphics[width=6.8cm]{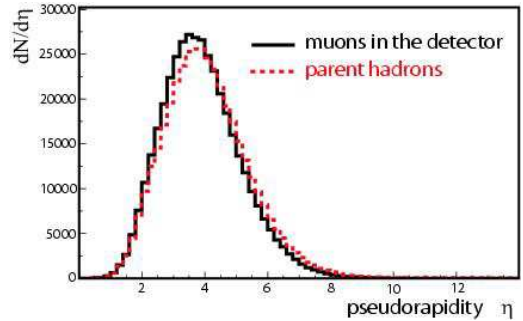}
\vspace*{-0.8cm}
\caption{Pseudorapidity distribution predicted by CORSIKA/QGSJET
simulations for muons detectable in KASCADE and for the parent hadrons
of the same muons.}
\label{rapid}
\vspace*{-0.5cm}
\end{figure}

Another approach to test the hadronic interaction models via the muon
component is possible due to the excellent angular resolution of the 
KASCADE Muon Tracking Detector ($\approx$0.35$^\circ$).
Measuring the relative angles $\tau$ and $\rho$ between 
single shower muons and the 
shower axis the pseudorapidity 
$\eta = ln{\frac{2\,p_\parallel}{p_t}} 
\approx  -ln{(\sqrt{\tau^2+\rho^2}/2)}$ can be calculated.
Fig.~\ref{rapid} shows the correlation of the pseudorapidity of
the muons measurable in the MTD and of the parent hadrons~\cite{zabi}. 
The rapidity distribution of hadrons generated in 
high-energy hadronic 
interactions is still an open question and an important parameter
for the model building.

\subsection{Tests of the shower development}
Investigating further shower observables at KASCADE
like the muon arrival 
time distributions~\cite{times},
the electron and muon lateral distributions or the muon production
heights~\cite{hei04} enables more to scrutinize the 
shower development itself rather than the hadronic interactions. 
In summary, within the given sensitivity of KASCADE
CORSIKA describes the shower development reasonably well and no 
significant deviations were found.  

\section{Conclusions and outlook}

Measurements with the multi-detector setup KASCADE provide  
plenty of high quality data to investigate the physics of the knee
in the cosmic ray energy spectrum. By analyzing these data no 
significant large scale anisotropy is found in the energy region
of the knee and no point source candidates could be identified, 
neither for charged cosmic rays nor for a sample with enhanced number of 
gamma-ray candidates. Concerning energy reconstruction,    
conclusive evidence has been given that the knee is caused by a
decrease of the flux of light primaries, where the knee positions
show a dependence on the primary mass group.
Systematic uncertainties for the estimate of the elemental
composition are dominated by the inadequacy of 
the hadronic interaction models underlying 
the reconstruction of energy spectra of single mass groups. Hence,
still there are only weak constraints for detailed
astrophysical models to explain the knee in the
primary cosmic ray energy spectrum. 
 
Indications for the inadequate description of the 
hadronic interactions at the atmosphere are also 
given by additional KASCADE data analyses taking the advantage of 
the multi-detector information, i.e.~investigations of the hadron 
component in air-showers or of muon properties measured 
for different muon energy thresholds. 
These investigations of observable correlations 
have shown that none of the present hadronic
interaction models is able to describe all the KASCADE data
consistently (on a level of a few percent).
Recently some efforts are made to
sample the information from accelerator experiments and cosmic ray 
investigations~\cite{engel} to improve the hadronic interaction
models. 

The multi-detector concept of the KASCADE experiment has been 
translated to 
its extension KASCADE-Grande~\cite{kg-haungs}, 
accessing higher primary energies up to $10^{18}\,$eV
which will prove the existence of a 
knee-like structure for heavy elements. 
In future, by having the data of the KASCADE-Grande experiment 
and by further improving the hadronic interaction models better
constraints especially at higher primary energies are expected.
Thus cosmic ray physics at energies around the knee remains a 
vital field of research with high scientific interest.

\vspace*{0.2cm}
{\small
\noindent {\bf Acknowledgment}: KASCADE is supported by 
the Ministry for Research and Education of Germany, 
the Polish State Committee for Scientific Research 
(KBN grant for 2004-06) and the Romanian National Academy 
for Science, Research and Technology.
}

\end{document}